\documentclass[iop,apj,numberedappendix,appendixfloats]{emulateapj}
\usepackage{ulem}
\usepackage{xcolor}
\usepackage{natbib} 
\usepackage{multirow}
\usepackage{amsmath}
\usepackage{url}
\usepackage{enumerate}

\shorttitle{Wideband Monitoring Observations of PSR~J1803$-$3002A}
\shortauthors{Lei Zhang et al.}

\begin{document}
\title{Wideband Monitoring Observations of PSR~J1803$-$3002A in the Globular Cluster NGC\,6522}

%\correspondingauthor{Lei Zhang}
\email{$^*$leizhang996@nao.ac.cn, dili@nao.ac.cn}

\author{Lei Zhang$^{1,2*,3}$, Richard N. Manchester$^{3}$, Andrew D. Cameron$^{3}$,  George Hobbs$^{3}$, Di Li$^{2*,5,11}$, Shi Dai$^{3}$, Qijun Zhi$^{4,6}$, Zonghong Zhu$^{1}$, Jingbo Wang$^{7,8,9}$, Lawrence Toomey$^{2}$, Yi Feng$^{2,11,3}$, Shuangqiang Wang$^{7,2}$, Songbo Zhang$^{10}$}
\affiliation{
$^1$ School of Physics and Technology, Wuhan University, Wuhan 430072, China\\
$^2$ National Astronomical Observatories, Chinese Academy of Sciences, A20 Datun Road, Chaoyang District, Beijing 100101, China\\
$^3$ CSIRO Astronomy and Space Science, PO Box 76, Epping, NSW 1710, Australia\\
$^4$ Guizhou Provincial Key Laboratory of Radio Astronomy and Data Processing, Guizhou Normal University, Guiyang 550001, China\\
$^5$ NAOC-UKZN Computational Astrophysics Centre, University of KwaZulu-Natal, Durban 4000, South Africa\\
$^6$ School of Physics and Electronic Science, Guizhou Normal University, Guiyang 550001, China\\
$^7$ Xinjiang Astronomical Observatory, 150, Science-1 Street, Urumqi 830011, China\\
$^8$ Key Laboratory of Radio Astronomy, Chinese Academy of Sciences, 150 Science 1-Street, Urumqi, Xinjiang, 830011, People's Republic of China \\
$^9$ Xinjiang Key Laboratory of Radio Astrophysics, 150 Science 1-Street, Urumqi, Xinjiang, 830011, People's Republic of China\\
$^{10}$ Purple Mountain Observatory, Chinese Academy of Sciences, Nanjing 210008, China
$^{11}$University of Chinese Academy of Sciences, Beijing 100049, China.
}
%{\small Received ; accepted }
  
%\adc{\sout{which is}} 

\begin{abstract}
We report the first wideband monitoring observations of PSR~J1803$-$3002A, a relatively bright millisecond pulsar in the globular cluster NGC\,6522 with a spin period of 7.1\,ms and no known binary companion. These observations were performed using the Parkes 64-m radio telescope with the Ultra-Wideband Low (UWL) receiver system, which covers  704 to 4032\,MHz. We confirm that PSR~J1803$-$3002A is an isolated millisecond pulsar located near the cluster center and probe the emission properties of the pulsar over the wide observed band. The mean pulse profile consists of three components, with the outer components becoming more prominent at higher frequencies, and a mean spectral index for the pulsed emission of $-1.66\pm0.07$ over the observed band. The fractional linear and circular polarization increase with increasing frequency, which is unusual for pulsars. We determine a Faraday rotation measure of $-107\pm 6$~rad~m$^{-2}$ for the pulsar. PSR~J1803$-$3002A is a distant pulsar in the Galactic plane, but our observations show no evidence of pulse broadening due to interstellar scattering. These results demonstrate the power of ultra-wideband receivers and signal processing systems.
\end{abstract}

\section{Introduction}\label{sec:intro}
Millisecond pulsars (MSPs) in globular clusters (GCs) are important diagnostic tools for a suite of astrophysical problems. GC MSPs can be used to study the evolution of binary systems \citep[e.g.,][]{Ginzburg20}, the gravitational potential of the GC \citep[e.g.,][]{Freire17}, the dynamical interactions in the GC core \citep[e.g.,][]{Ye19}, the interstellar medium (ISM) and its magnetization \citep[e.g.,][]{Abbate20}, the intracluster medium, e.g., in 47 Tucanae, \citep{Abbate18} and more generally \citep{Naiman20}, the equation of state of nuclear matter \citep{of16} and neutron star retention in GCs \citep[e.g.,][]{Pfahl02}.

Radio searches for and/or studies of the pulsars in GCs can be challenging, because many of them are at large distances, which makes their flux density typically  weak and signals strongly distorted by propagation through the ISM. Moreover, the pulsars in GCs are often members of tight binary systems, causing large changes in their observed spin period and sometimes periodic eclipsing of the radio signal.

Baade's Window is an area of the sky close to the Galactic Center direction with relatively low amounts of interstellar dust and gas along the line of sight from the Earth. The globular cluster NGC\,6522 lies near the center of Baade's Window and, with an estimated distance of 7.7~kpc \citep{har10a}\footnote{http://physwww.mcmaster.ca/$\sim$harris/Databases.html}, is relatively close to the Galactic Center. This GC is well studied and, with a stellar age of $>$12\,Gyr, is possibly the oldest GC in our Galaxy \citep{Barbuy09}. It is a core-collapsed cluster \citep{kno+18} and searching for pulsars in such clusters has the potential to identify exotic systems. Although the majority of the pulsars  are expected to be isolated because of the high interaction rate \citep[e.g.,][]{Verbunt14}, eccentric MSP binaries with massive companions (neutron stars as well as white dwarfs) can be produced through exchange encounters, e.g.\ B2127$+$11C in M15, J1807$-$2500B in NGC 6544, and J1835$-$3259A in NGC 6652 \citep[][]{Prince91,Lynch12,DeCesar15}. Another such system, J0514-4002A, is found in NGC 1851, which possesses a massive and dense core \citep[][]{Ridolfi19}, although not a core-collapsed cluster. The ultimate reward of such searches would be a MSP -- black-hole binary, which may result from similar conditions.

Three MSPs (PSRs~J1803$-$3002A, J1803$-$3002B and J1803$-$3002C) have been found in the cluster by previous pulsar surveys. PSR~J1803$-$3002A is relatively strong and was discovered in the Parkes Globular Cluster survey at 20\,cm \citep{Possenti05}. The other two pulsars were discovered using the Green Bank Telescope \citep[][private communication, P.\,C.\,C.~Freire]{Freire08}. To date, there have been no published long-term timing measurements, polarization observations or flux density measurements for the pulsars in NGC 6522. Their approximate positions, pulse periods, and dispersion measures (DMs) were obtained from pulsar-search data.

We have carried out the first wide-bandwidth observations, timing and polarization analyses for pulsars in NGC\,6522 using the Ultra-Wideband Low (UWL) receiver system installed on the Parkes 64m radio telescope \citep{Hobbs20}. In Section~\ref{sec:obs}, we describe the details of our observations and processing methods. We then present a wideband timing solution for PSR~J1803$-$3002A in Section~\ref{sec:timing}, pulse polarization profiles and rotation measure (RM) in Section~\ref{sec:prof}, and flux densities and spectral properties in Section~\ref{sec:spectra}. In  Section~\ref{sec:search} we discuss our search results for other pulsars in the direction of NGC\,6522. A  summary is provided in Section~\ref{sec:summary}.

%%%%%%%%%%%%%%%%%%%%%%%%%%%%%%%%%%%%%%%%%%%%%%%%%%%%%%%%%%%%%%%%%%%%%%%%%%%%%%%%%
\section{Observations and Processing}\label{sec:obs}
We have carried out 17 observations of PSR~J1802$-$3002A in NGC\,6522 between 2019 August 4 (MJD 58669.28) and 2020 July 4 (MJD 59034.69) using the UWL receiver system on the Parkes 64-m radio telescope. The telescope was pointed at the nominal cluster center: 18$^{\rm h}\;03^{\rm m}\;34\fs02$, $-30\degr\;02\arcmin\;02\farcs3$ \citep{har10a}. Table~\ref{tb:NGC6522A_obs} lists each observation, the start time of the observation in Coordinated Universal Time (UTC) and modified Julian date (MJD), the observation length, the observing mode (fold or search), and the project ID.
 
For the search-mode observations, total intensity data were recorded with 2-bit sampling every 64\,$\mu s$ in each of the 1\,MHz-wide frequency channels (3328 channels across the whole band from 704\,MHz to 4032\,MHz). Within each 1\,MHz channel, the data were coherently de-dispersed at a DM of 192.37\,pc cm$^{-3}$ (corresponding to the mean of the values of the known pulsars in the GC). We carried out a periodicity search for pulsars in the GC within a DM range of 182--202\,pc cm$^{-3}$ using a relatively wide band (970--3018\,MHz) which was cleaned of radio frequency interference (RFI). We also carried out a periodicity search for pulsars in the GC direction with a DM range of 0--1000\,pc cm$^{-3}$ using a narrower RFI-clean band: 1216--1472\,MHz. For the periodicity search, we split each search mode observation into 4400\,s blocks and carried out searches for pulsar signals in the Fourier domain using the \textsc{presto}\footnote{http://www.cv.nrao.edu/sransom/presto/} software suite \citep{Ransom02}, with a Fourier drift-rate $z$ range \citep[see][for the definition of $z$]{Andersen18} of $\pm 200$ to give sensitivity to pulsars in short-period binary orbits \citep{Ng15}. Pulsars with narrow pulse profiles have many significant harmonics of the spin frequency in the power spectrum and so we summed up to 16 harmonics to increase the significance of the final detection. We were only able to detect PSR~J1803$-$3302A, which has a pulse period of $\sim$7.1\,ms, in any of our searches. 

We subsequently folded each search-mode observation  with 60\,s subintegrations and 256 phase bins using \textsc{dspsr} \citep{van11}. We removed 5\,MHz at each edge of each of the 26 sub-bands which comprise the full bandwidth of the UWL in order to avoid the effects of aliasing \citep[see more details in][]{Hobbs20} and then manually removed data affected by RFI in frequency and time for each channel and sub-integration. The folded search-mode data were helpful in building a timing solution for PSR~J1803$-$3302A.

In order to obtain a phase-connected timing solution for PSR J1803$-$3002A and accurately calibrate both its flux density and polarization properties, we carried out additional follow-up fold-mode observations that cover the entire band of the UWL receiver. For these fold-mode observations, the data was coherently de-dispersed at a DM of 192.37\,pc cm$^{-3}$ with full Stokes information recorded, channelized with 1\,MHz channels, reduced to 512 phase bins per period, and written to disk with 20\,s subintegrations. We also recorded a short (1 minute) observation of a switched calibration noise source \citep[see more details in][]{Hobbs20} before each observation in order to allow for polarisation calibration of each observation. To calibrate the polarization response of the UWL feed, we used multiple observations of the bright MSP PSR~J0437$-$4715 that covered a wide range of parallactic angles \citep{van04}, taken during the commissioning of UWL in 2018 November.

To get an initial timing solution of PSR J1803$-$3002A, we measured pulse times of arrival (ToAs) using programs from the \textsc{psrchive} package \citep{Hotan04} to sum each observation in both frequency and polarisation, before partially summing in time to form 20-min sub-integrations and cross-correlating each summed profile against a standard reference pulse profile. The \textsc{tempo2} software package \citep{Hobbs06} was then used to derive an initial timing solution for the pulsar from the ToAs. We then re-fitted each observation by using this initial timing solution to generate more precise ToAs, which in turn enabled us to iteratively obtain more precise timing solutions. For a final solution, the data were again partially summed in time to form 20-min sub-integrations and in frequency to form 11 unequal sub-bands defined by three groups: five equal sub-bands from 704 to 1344\,MHz (RF Band 1) four sub-bands from 1344 to 2368\,MHz (RF Band 2), and two sub-bands from 2368 to 4032 MHz (RF Band 3). ToAs were then formed using different noise-free templates for the three RF bands obtained by fitting to mean pulse profiles for these bands using the program {\sc paas}. The final phase-coherent timing solution was then obtained using \textsc{tempo2} with jumps between the RF bands to allow for the different templates used.

To probe the pulsed emission properties of PSR~J1803$-$3002A, we again used programs from the {\sc psrchive} package. We constructed average polarization profiles by summing the pulsar's calibrated observations in time using the timing solution of the pulsar to ensure phase alignment. To obtain a high signal-to-noise ratio (S/N) polarization profile for further analysis, we selected the average polarization  profiles of the pulsar in the three RF bands for each observation for which the S/N was more than 10. They were then included in the summation using \textsc{psrwt}. Since the pulse profile has no measurable linear polarisation in RF Band 1 (see Section~\ref{sec:prof} below) the rotation measure (RM) was determined across the two higher-frequency bands using the \textsc{rmfit} program. The measured RM was then used to refer all measured position angles (PAs) to the overall band center, 2688\,MHz, before summing in frequency to form the average polarization profiles. The pulse widths at $50\%$ (W$_{50}$) and $10\%$ (W$_{10}$) of the pulse peak were measured using the program \textsc{pdv} from noise-free profiles obtained using the program  \textsc{paas}.
 
Flux densities were also measured from the summed, calibrated observations using the program \textsc{pdv}, summing the data across phase bins as appropriate. The flux density scale was set using observations of the radio quasar 0407-658 assuming a flux density of 14.4\,Jy at 1400\,MHz with a power-law spectral index of $-1.189$. A separate flux density measurement was made for each of the 26 sub-bands comprising the whole bandwidth from (704 to 4032\,MHz) of the UWL receiver. These sub-band flux densities were then used to derive the spectral properties (see Section~\ref{sec:spectra} below).
 
Conditional on data embargoes\footnote{Typically 18 months following the observation.}, the data from these observations are available from the CSIRO's data archive\footnote{https://data.csiro.au/} \citep{Hobbs11}. We have also produced a publicly downloadable data collection containing our processed data files\footnote{ https://doi.org/10.25919/5f45d801827d6} \citep{Zhang20}. This data collection contains the integrated and RFI-removed file for each observation, and the adopted timing model for PSR~J1803$-$3002A.

\begin{table}[]
\footnotesize
\caption{Observations of PSR~J1803$-$3002A}\label{tb:NGC6522A_obs}
\begin{tabular}{cccccc}
\hline
No. & Start        & Date     & Duration & Obs.mode & ID  \\
    & (UTC)        & (MJD)    & (second) &          &           \\ \hline
1 & 2019 Aug. 4  & 58699.28 & 33880 & search & P1006\\
2 & 2019 Nov. 22 & 58809.22 & 12980 & search & PX501\\
3 & 2019 Nov. 23 & 58810.05 & 26840 & search & PX501\\
4 & 2019 Nov. 24 & 58811.05 & 26840 & search & PX500\\
5 & 2019 Nov. 29 & 58816.30 & 4591  & fold   & P456 \\
6 & 2020 Jan. 27 & 58875.93 & 7216  & fold   & PX500\\
7 & 2020 Feb. 12 & 58891.75 & 10511 & fold & P1006\\
8 & 2020 Feb. 15 & 58894.88 & 10816 & fold & P1006\\
9 & 2020 Feb. 25 & 58904.84 & 2185  & fold & P1006\\
10 & 2020 Apr. 06 & 58945.88 & 2752 & fold & P1006\\
11 & 2020 Apr. 14 & 58953.83 & 6842 & fold & P1006\\
12 & 2020 Apr. 15 & 58954.85 & 9294 & fold & P1006\\
13 & 2020 May. 05 & 58974.74 & 3015 & fold & PX501\\
14 & 2020 May. 16 & 58985.79 & 4664 & fold & P1006\\
15 & 2020 May. 26 & 58995.71 & 4496 & fold & P1006\\
16 & 2020 Jun. 14 & 59014.78 & 2364 & fold & P1006\\
17 & 2020 Jul. 04 & 59034.69 & 4631 & fold  & P1063\\
\hline
\end{tabular}
\end{table}

%%%%%%%%%%%%%%%%%%%%%%%%%%%%%%%%%%%%%%%%%%%%%%%%%%%%%%%%%%%%%%%%
\section{Results and Discussion}\label{sec:results}
\subsection{Timing Solution}\label{sec:timing}

\begin{figure*}[]
\centering
\includegraphics[width=1.0\linewidth]{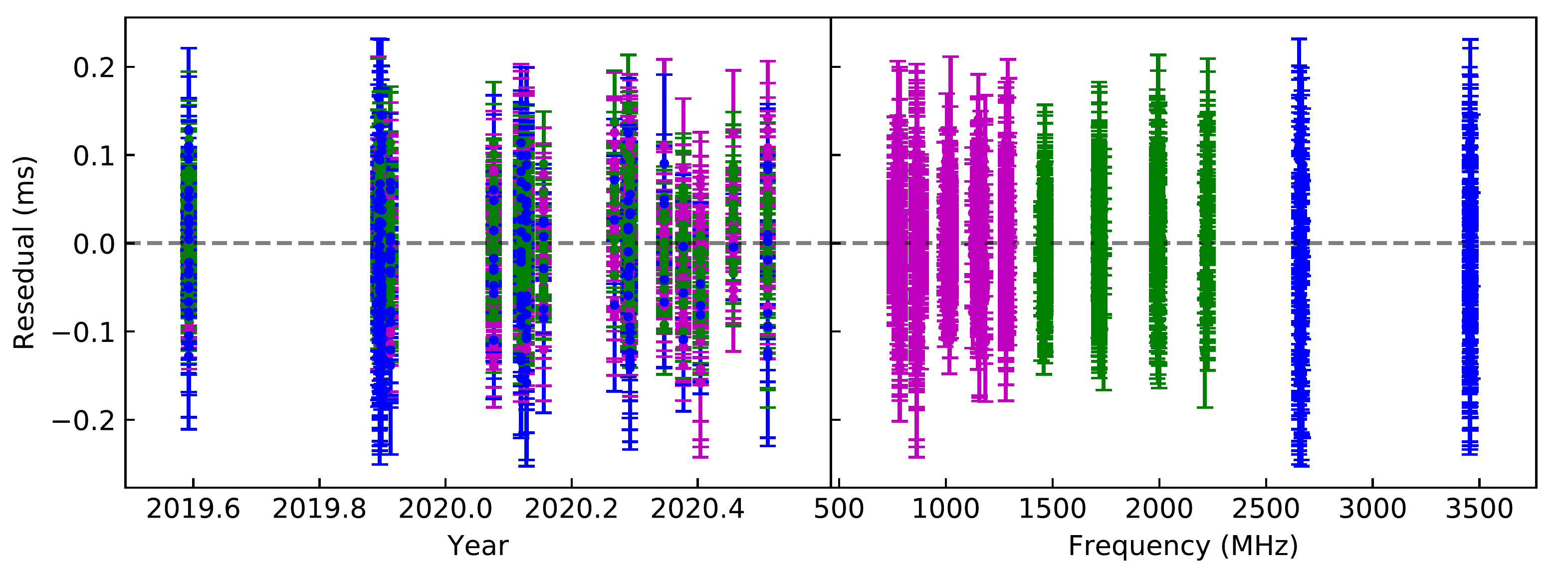}\\
\caption{Timing residuals for PSR~J18003$-$3002A based on the Parkes UWL monitoring observations over the 704 to 4032\,MHz UWL band. The dashed line marks the zero-line and corresponds to the arrival times predicted by the timing solution. The different colours for the residuals represent the three RF bands:  Band 1 (704--1344\,MHz; purple), Band 2 (1344--2368\,MHz; green) and Band 3 (2368--4032\,MHz; blue).}
\label{fig.NGC6522A_res}
\end{figure*}

We used the \textsc{tempo2} software package \citep{Hobbs06} with the DE436 solar system ephemeris and the TT(TAI) time standard to obtain a phase-connected timing solution extending over 335\,days. The timing residuals for PSR~J18003$-$3002A are shown in Figure~\ref{fig.NGC6522A_res}. Residuals have an approximately Gaussian distribution about the zero line, with no trends in either time or frequency, indicating that there is neither detectable red timing noise nor non-dispersive frequency dependence in the ToAs. Although the data span is slightly less than one year, the pulsar position is well determined.

The timing solution of the pulsar is presented in Table~\ref{tb:NGC6522A_par}. As discussed in Section~\ref{sec:obs}, since the pulse profile evolves significantly with frequency (see Section~\ref{sec:prof} below), we use different profile templates for the three RF bands and fit for jumps between the bands, with Band 2 as reference. Timing residuals are shown in Figure~\ref{fig.NGC6522A_res} with the different RF bands identified by color. We also present derived parameters in  Table~\ref{tb:NGC6522A_par}. According to our timing solution, the pulsar's position is offset from the nominal cluster centre by $14\farcs5$ which, for a cluster distance of 7.7\,kpc, corresponds to a perpendicular offset of 0.54~pc.

\begin{table}[]
\centering
\caption{Parameters for PSR~J1803$-$3002A}\label{tb:NGC6522A_par}
\begin{tabular}{llll}
\hline
\multicolumn{4}{c}{Timing parameters}  \\
\hline 
\multicolumn{2}{l}{RA (J2000)}          &\multicolumn{2}{l}{18$^{\rm h}\;03^{\rm m}\;35\fs1290(1)$}          \\
\multicolumn{2}{l}{Dec. (J2000)}        & \multicolumn{2}{l}{$-30\degr\;02\arcmin\;00\farcs05(2)$}          \\
\multicolumn{2}{l}{Spin frequency, $\nu$ ($\rm s^{-1}$)}    & \multicolumn{2}{l}{140.8174521908(3)}       \\
\multicolumn{2}{l}{Spin frequency derivative, $\dot{\nu}$ ($\rm s^{-2}$)} &\multicolumn{2}{l}{ $-7.93(19)\times10^{-16}$}  \\
\multicolumn{2}{l}{Dispersion measure, DM (cm$^{-3}$ pc)}       &\multicolumn{2}{l}{ 192.6234(5)}  \\
\multicolumn{2}{l}{Timing epoch (MJD)}        &\multicolumn{2}{l}{ 58700}                     \\
\multicolumn{2}{l}{Date of first observation (MJD)}         & \multicolumn{2}{l}{58699.28 }          \\
\multicolumn{2}{l}{Date of last observation (MJD)}           &\multicolumn{2}{l}{ 59035.76}              \\
\multicolumn{2}{l}{Total time span (days)}                        &\multicolumn{2}{l}{ 335.48 }           \\
\multicolumn{2}{l}{Time standard}                                  & \multicolumn{2}{l}{TT(TAI)} \\
\multicolumn{2}{l}{Time units  }                                & \multicolumn{2}{l}{TCB} \\
\multicolumn{2}{l}{Solar-system ephemeris}                      & \multicolumn{2}{l}{DE436}  \\
\multicolumn{2}{l}{Weighted RMS residual ($\mu$s)}   &\multicolumn{2}{l}{ 38.816}  \\
\multicolumn{2}{l}{Reduced $\chi^{2}$}                      & \multicolumn{2}{l}{1.16 } \\
\hline
\multicolumn{4}{c}{Derived parameters}  \\
\hline
\multicolumn{2}{l}{Galactic longitude ($\degr$)}                 &\multicolumn{2}{l}{ 1.027} \\
\multicolumn{2}{l}{Galactic latitude ($^{\circ}$)}          &\multicolumn{2}{l}{ $-$3.929 } \\
\multicolumn{2}{l}{Spin period, P (s) }                 &\multicolumn{2}{l}{ 0.00710139250812(1)}   \\
\multicolumn{2}{l}{Spin period derivative, $\dot{\rm P}$}   &\multicolumn{2}{l}{ 4.0(1) $\times$ 10$^{-20}$}\\
\multicolumn{2}{l}{Characteristic age, $\tau _{c}$ (Myr)} & \multicolumn{2}{l}{2814}    \\
\multicolumn{2}{l}{Surface magnetic field, $B_{\rm surf}$ (G)}  & \multicolumn{2}{l}{5.39 $\times$ 10$^{8}$}  \\
\hline                                              \multicolumn{4}{c}{Profile and Polarization parameters}                     \\
\hline 
\multicolumn{2}{l}{Frequency} & W$_{50}$ (\degr)   & W$_{10}$ (\degr)   \\  
\hline
\multicolumn{2}{l}{1024 MHz}         & 23.9           & 74.0            \\
\multicolumn{2}{l}{1856 MHz}         & 22.8          & 146                  \\
\multicolumn{2}{l}{3200 MHz}         & 66.3           &   169          \\
\hline
\multicolumn{2}{l}{Rotation measure (rad m$^{-2}$) }      & $-$107(6)        \\
\hline
\multicolumn{4}{c}{Flux densities (mJy)}            \\
\hline
S$_{768}$                    &    1.89(30)      & S$_{2432}$                   &    0.38(13)  \\
S$_{896}$                    &    1.85(6)      & S$_{2560}$                   &    0.28(4)   \\
S$_{1024}$                   &    1.47(8)      & S$_{2688}$                   &    0.22(13)   \\
S$_{1152}$                   &    0.95(3)      & S$_{2816}$                   &    0.26(7)   \\
S$_{1280}$                   &    0.82(14)      & S$_{2944}$                   &    0.21(3)   \\
S$_{1408}$                   &    0.70(5)      & S$_{3072}$                   &    0.31(12)   \\
S$_{1536}$                   &    0.60(1)      & S$_{3200}$                   &    0.20(1)   \\
S$_{1664}$                   &    0.56(2)      & S$_{3328}$                   &    0.23(5)   \\
S$_{1792}$                   &    0.45(4)      & S$_{3456}$                   &    0.22(2)    \\
S$_{1920}$                   &    0.37(4)      & S$_{3584}$                   &    0.23(2)   \\
S$_{2048}$                   &    0.31(4)      & S$_{3712}$                   &    0.18(3)   \\
S$_{2176}$                   &    0.38(3)      & S$_{3840}$                   &    0.25(13)    \\
S$_{2304}$                   &    0.34(14)      & S$_{3968}$                   &    0.20(11)   \\
\hline
\multicolumn{4}{c}{Spectral parameters} \\
\multicolumn{2}{l}{Power-law}       & \multicolumn{2}{l}{Log-parabolic} \\
\hline
$\alpha$ & $-$1.66(7) & $a$  & 1.13(19) \\      
$\beta$  & 1.28(5)    & $b$  & $-$2.17(9)    \\
&  &  $c$  & 0.15(1)  \\
\multicolumn{2}{l}{Reference frequency ($\nu_0$) } & 1000\,MHz & \\
\hline
\multicolumn{4}{l}{{\bf Note:} Uncertainties in parentheses are $1\,\sigma$ and refer to the} \\
\multicolumn{4}{l}{last quoted digit.}
\end{tabular}
\end{table}

\subsection{Polarization profiles and RM}\label{sec:prof}
\begin{figure}[]
\centering
\includegraphics[ width=0.9\linewidth]{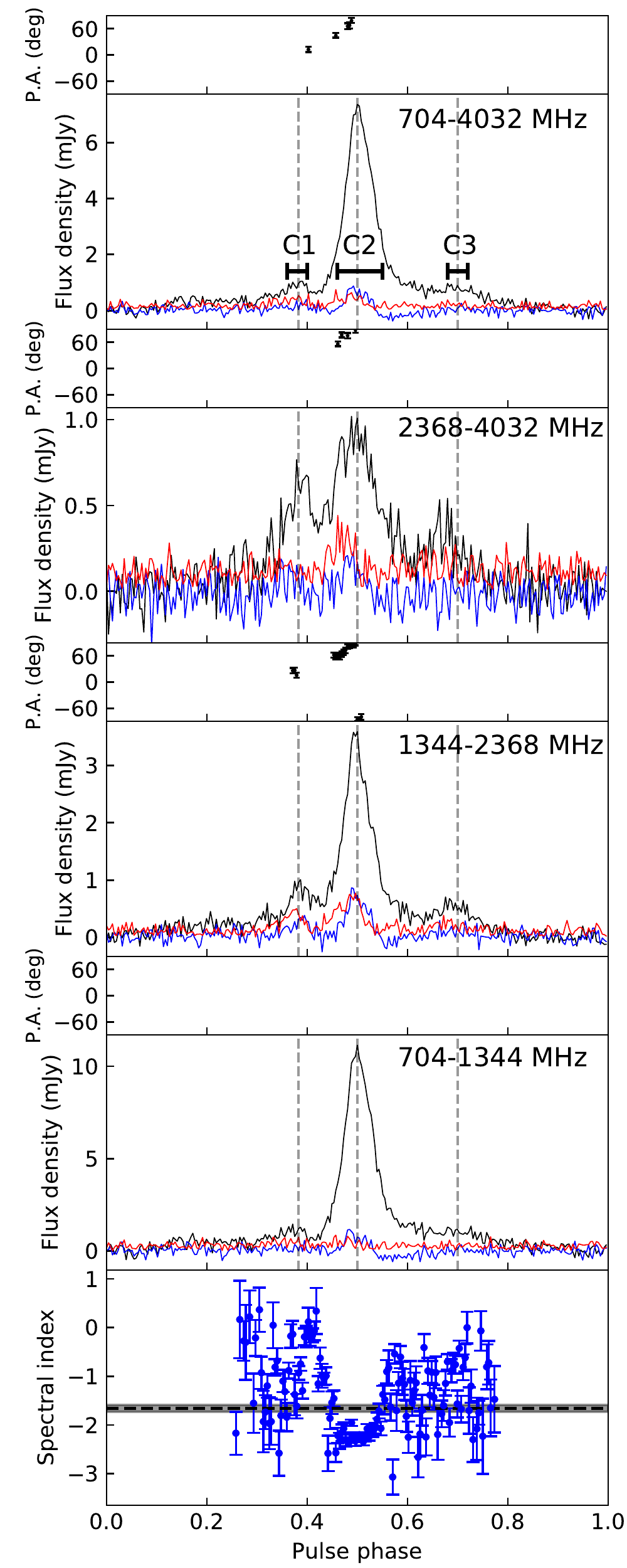}\\
\caption{Average polarization profiles for PSR~J1803$-$3002A in the three RF bands between 704 and 4032\,MHz and the sum over the whole band. The red line is the linear polarization profile, the blue line is the circular polarization profile and the black line is the mean flux density profile. Black dots in the upper panel give the linear position angle (PA) referred to the overall band center of the averaged profiles, 2688\,MHz. The three main pulse components are labeled on the whole-band profile. The bottom panel shows phase-resolved spectral index with uncertainties (blue points with error bars) for the pulsar. The horizontal dashed line marks the fitted power-law spectral index for the whole profile ($-$1.66) and the gray area shows its uncertainty ($\pm$0.07). The horizontal bars in the top panel mark the region of each component used to calculate its spectral index.}
\label{fig.NGC6522A_profs}
\end{figure}

In Figure~\ref{fig.NGC6522A_profs}, we present the average polarization profile for PSR~J1803$-$3002A for the entire UWL observation band centred at 2368\,MHz in the top panel. We also show the average polarization profiles of the pulsar in the three RF bands, centered at 1024, 1856, and 3200\,MHz, respectively. In all cases, the position angles are referred to the overall band center, 2688 MHz, using the derived RM (see below). The bottom panel of the figure shows the phase-resolved power-law spectral indices and their uncertainties.

%% Faraday rotation
Since searches over a wide range of RM detected no measurable linear polarization in RF Band 1, we determine the RM using data from just RF Bands 2 and 3. The derived RM is $-107\pm5$~rad~m$^{-2}$. We have not corrected this result for the ionospheric contribution, expected to be relatively small, between $-0.2$ and $-2$~rad~m$^{-2}$, at the latitude of Parkes \citep{hmvd18}. However, to allow for the additional uncertainty in the measured value, we increase quoted uncertainty to 6~rad~m$^{-2}$ as given in Table~\ref{tb:NGC6522A_par}.

With the wide-band and coherently de-dispersed observations, we are able to investigate changes of the observed pulse shape relating to intrinsic profile changes and propagation effects in the ISM over a wide frequency range. We measured the width of the profiles for PSR~J1803$-$3002A at 50\% and 10\% of the peak flux density in the three RF bands and have listed these results in Table~\ref{tb:NGC6522A_par}. We notice that there is a systematic increase in pulse width with increasing frequency. This is entirely due to the flatter spectrum of the outer components relative to the central component. Like many other wide-profile MSPs \citep[e.g. PSRs~J0437$-$4715 and J2124$-$3358,][]{Dai15}, the pulse components in PSR~J1803$-$3002A have a central pulse phase that is frequency-independent. In general, MSPs do not follow the radius-to-frequency mapping commonly inferred for normal pulsars \citep{Cordes78}.  For MSPs, the emission region may be close to the light cylinder and caustic effects are likely to be important in defining the observed pulse profile \citep{Ravi10}.

For PSR~J1803$-$3002A, the observed pulse profiles indicate three components (labeled as C1, C2 and C3 in the top panel of Figure~\ref{fig.NGC6522A_profs}), with the leading and trailing components becoming relatively stronger compared to the central component as the observing frequency increases. As shown by \citet{Dai15}, many MSPs show variations in relative spectral index between different profile components.

Linear polarization is commonly seen in MSPs \citep[e.g.,][]{Dai15} although the observed PA variations seldom follow a rotating-vector-model pattern \citep{Radhakrishnan69}. Compared to other MSPs, PSR~J1803$-$3002A has a relatively small amount of linear polarization. Furthermore, it is unusual in that the fractional linear polarisation decreases with decreasing frequency (Figure~\ref{fig.NGC6522A_profs}).

As mentioned above, no significant linear polarisation was observed for Band 1 data in a search over a wide range of RMs and variations in ionospheric RM are too small to depolarize the profile. Depolarisation can also arise in highly scattered pulsars due to propagation through turbulent plasma components that are irregularly magnetised.  For example, PSRs~B2111$+$46 \citep{Noutsos15} and J0742$-$2822 \citep{Xue19} show a rapid decrease in linear polarisation with decreasing frequency at frequencies $<300$\,MHz where there is significant scattering by the ISM. However, we have not seen any evidence for interstellar scattering even in the low-frequency band (scattering timescale $<$\,0.5\,ms). Consequently, the observed low polarisation cannot be attributed to stochastic Faraday rotation across the scattering disk.
We therefore conclude that the observed low degree of linear polarization for PSR~J1803$-$3002A at low frequencies is intrinsic.

\subsection{Flux densities and spectral properties}\label{sec:spectra}

\begin{figure}[]
\centering
\includegraphics[width=1.0\linewidth]{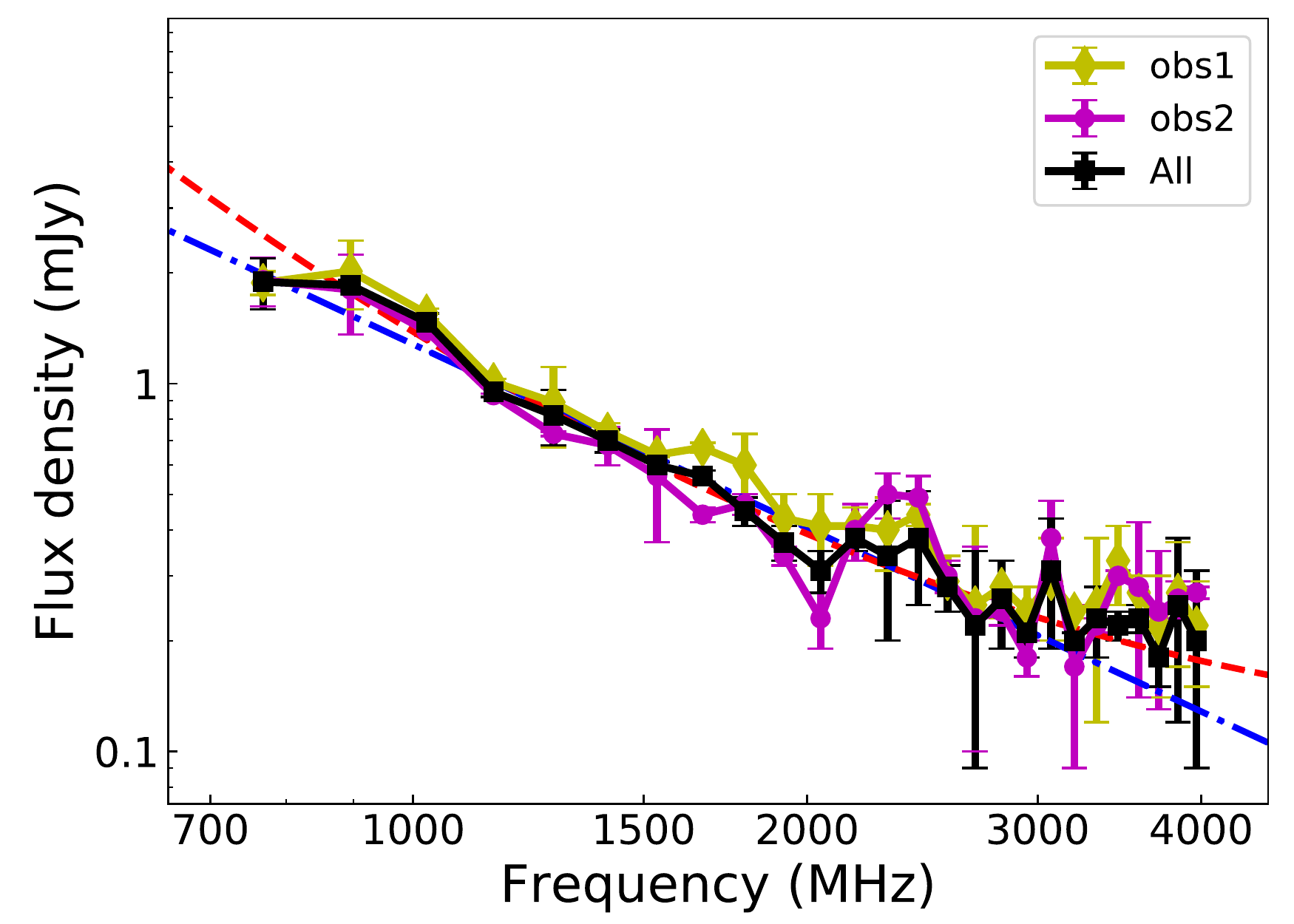}\\
\caption{Flux density as a function of frequency with 26 frequency sub-bands (each sub-band with 128\,MHz bandwidth) over the whole band of the UWL for PSR~J1803$-$3002A. The legend indicates the color/line scheme used to represent specific fold-mode observations or the average of all fold-mode observations. The dot-dashed and dashed lines represent the spectral fitting result of the average of all fold-mode observations for a power-law and LPS spectrum, respectively. See the text in Section~\ref{sec:spectra} for more details.}
\label{fig.NGC6522A_flux}
\end{figure}

We list the calibrated flux densities averaged over all observations for the  26 sub-bands (each with a 128 MHz bandwidth) across the UWL band in Table~\ref{tb:NGC6522A_par}. Figure~\ref{fig.NGC6522A_flux} shows the flux density values separately from observations taken from 2019 November to 2020 February (labeled ``obs1"), and from 2020 April to May (labeled ``obs2"), together with the average of all fold-mode observations (labeled ``All"). We first fit a simple power law:
\begin{equation}
\text{S}_{\nu}=\beta x^{\alpha},
\label{eq:PL}
\end{equation} 
where $x = {\nu}/{\nu_{0}}$, $\alpha$ is the spectral index and $\beta$ is a scaling constant. The results of the power-law fit are given in Table~\ref{tb:NGC6522A_par} and shown in Figure~\ref{fig.NGC6522A_flux}. The fitted spectral index is $-1.66\pm0.07$. Figure~\ref{fig.NGC6522A_flux} suggests that there may be some curvature in the spectrum although it is of marginal significance. We have also fitted the measured flux densities with a log-parabolic spectrum (LPS):
\begin{equation}
\log_{10}\text{S}_{\nu} = ay^{2}+by+c,
\label{eq:cubic}
\end{equation}
where $y=\log_{10}({\nu}/{\nu_{0}})$, $a$ is the curvature parameter, $b$ is an effective spectral index and $c$ is a constant. The LPS fit result is presented in Table~\ref{tb:NGC6522A_par} and shown in Figure ~\ref{fig.NGC6522A_flux}. Overall, we believe the simple power-law fit to be a more reliable representation of the spectral properties of PSR J1803$-$3002A.

The ATNF Pulsar Catalogue\footnote{https://www.atnf.csiro.au/research/pulsar/psrcat/} \citep{Manchester05} lists spectral indices for 55 MSPs and these lie in the range $-$1.1 to $-$3.8, with a mean value of $-2.0$. The measured spectral index  for PSR~J1802$-$3002A, $-1.66$, is well within this range, and somewhat flatter than the mean value.

The bottom panel in Figure~\ref{fig.NGC6522A_profs} shows the phase-resolved power-law spectral index across the PSR J1803$-$3002A profile, clearly showing  variations related to the pulse components. We calculated the mean spectral indices for the central region of each component where the spectral index is relatively phase-independent to avoid the effects of overlap of the different components. The derived spectral indices are $-0.79\pm0.10$, $-2.20\pm0.05$ and $-1.01\pm0.13$ for the leading (C1), central (C2) and trailing (C3) components, respectively. Steep-spectrum central or ``core" components and flatter-spectrum outer or ``conal" components  are common in normal pulsars \citep{Backer76, Rankin83, Lyne88} and are usually interpreted in the context of the magnetic-pole model. In general, MSP profiles are more complex and do not have ``core-conal" structure, but such structures are seen in a few MSPs, e.g., PSR J1730$-$2304 \citep{Dai15}. This pulsar is similar to PSR J1803$-$3002A in having a central component with a steeper spectrum.

\subsection{Search for pulsars}\label{sec:search}

Globular clusters have proved to be fruitful for pulsar searches with 157 radio pulsars in 30 GCs discovered to date \citep{Manchester05,Freire08}. Almost all of these pulsars are MSPs and 85 are known to be in binary systems. A Fourier-domain implementation of the acceleration search in PRESTO \citep{Ransom02} has been use to find the vast majority of the binary pulsars in GCs. Such searches are effective when  $T \lesssim P_b/10$, where $T$ and $P_b$ are the observation duration and the binary-system orbital period, respectively. On the other hand, the sensitivity to fainter pulsars improves as  $\sqrt{T}$. As described in Section~\ref{sec:obs}, we split long search-mode observations into 4400-second blocks as a compromise between sensitivity and  computational feasibility giving sensitivity to binary systems with orbital period greater than about 0.5~d.  The nominal ($8\sigma$) sensitivity of our searches to a 4\,ms pulsar with DM $\sim$ 192\,pc cm$^{-3}$ and spectral index about $-$2.0 is 0.03 to 0.04\,mJy at 1400\,MHz.

Only PSR~J1803$-$3002A was detected in our pulsar search. Specifically, we were not able to detect either of two previously reported pulsars in the cluster (PSRs~J1803$-$3002B and J1803$-$3002C), nor any other pulsar lying within the telescope beam. Even folding our search data at the known period and DM of PSRs~J1803$-$3002B and J1803$-$3002C \citep{Freire08} gave no significant detection of either pulsar. As mentioned in the Introduction, there are no published flux densities for any of the three known pulsars in NGC~6522.

%%%%%%%%%%%%%%%%%%%%%%%%%%%%%%%%%%%%%%%%%%%%%%%%%%%%%%%%%%%%%%%%
\section{Summary}\label{sec:summary}

In this work, we present the previously unpublished timing solution (Section~\ref{sec:timing}) and emission properties (Sections~\ref{sec:prof} and \ref{sec:spectra}) of PSR~J1803$-$3002A in NGC 6522. A search for pulsars in the direction of NGC 6522 detected only PSR~J1803$-$3002A, confirming that it is the strongest of the three known pulsars in the cluster (Section~\ref{sec:search}). The mean pulse profile of PSR~J1803$-$3002A is rather weakly polarized and no linear polarization was detectable in the lowest RF band (704--1344\,MHz)  -- the pulsar RM was measured using data from the two higher RF bands. The overall radio spectral index for PSR~J1803$-$3002A, $-1.66\pm0.07$, is somewhat flatter than $-2.0$, the mean spectral index for MSPs. The three identifiable components of the pulse profile have significantly different spectral indices, with the central component having a steeper spectrum than the mean and the outer components having flatter spectra. We will continue to observe this pulsar with the UWL receiver to improve its timing parameters and to build up higher-S/N pulse polarization profiles.

Searches for pulsars in Baade's Window with the UWL receiver are on-going and will help us to probe the poorly understood distribution of distant pulsars. The observations described here were in either search mode or fold mode. The facility for simultaneous pulsar search as well as fold-mode observations and  HI spectroscopy \citep[cf.][]{li18} with the UWL system has been recently implemented. Furthermore, simultaneous folding at multiple pulsar periods is planned. These improvements will substantially enhance the observational efficiency for searches, confirmation, and timing.

%%%%%%%%%%%%%%%%%%%%%%%%%%%%%%%%%%%%%%%%%%%%%%%%%%%
\acknowledgments
\begin{acknowledgements}
This work is supported by National Key R\&D Program of China No. 2017YFA0402600, the National Natural Science Foundation of China (Grant No. 11988101, 11725313, 11690024, 11743002, 11873067), the Strategic Priority Research Program of the Chinese Academy of Sciences Grant No. XDB23000000 and the Foundation of Guizhou Provincial Education Department (No. KY(2020)003).
QJZ is supported by the National Natural Science Foundation of China (U1731218) and the Science and Technology Fund of Guizhou Province ((2016)-4008, (2017)5726-37).
JBW is supported by the Youth Innovation Promotion Association of Chinese Academy of Sciences.

The Parkes radio telescope is part of the Australia Telescope National Facility which is funded by the Australian Government for operation as a National Facility managed by CSIRO. We thank the Parkes team for their great effort to install and commission the UWL receiver system. This paper includes archived data obtained through the CSIRO Data Access Portal$^4$.
\end{acknowledgements}

%%%%%%%%%%%%%%%%%%%%%%%%%%%%%%%%%%%%%%%%%%%%%%%%%%%

\end{document}